\documentclass[twoside,12pt]{report}
\usepackage{graphicx}
\usepackage{amssymb, amsmath, amsxtra}
\usepackage{epsfig,subfigure}

\usepackage{latexsym}
\usepackage{bm}

 \usepackage{ntheorem}
           \theoremstyle{plain}
                      {\theorembodyfont{\rmfamily}
                      \theoremseparator{.}
                       
           \newtheorem{thm}{Theorem}[section]
           \theoremstyle{plain}
           \newtheorem{cor}{Corollary}[section]
           \theoremstyle{plain} \newtheorem{lem}{Lemma}[section]
           \theoremstyle{plain}
           \newtheorem{defn}{Definition}[section]
            
           \theoremstyle{plain}
           \newtheorem{rem}{Remark}[section]
            \newtheorem{prop}{Proposition}[section]}

\begin{document}
\begin{center}
{\Large \textbf{GROUP ANALYSIS OF NONLINEAR}\\[1ex] \textbf{INTERNAL WAVES IN
OCEANS}}\\[1.5ex]
 {\Large \textbf{I: Self-adjointness, conservation laws,}\\[.5ex] \textbf{invariant solutions}}\footnote{Published
 in \textit{Archives of ALGA}, vol. 6, 2009, pp. 19-44.}\\[1ex]
 {\sc Nail H. Ibragimov}\\
 Department of Mathematics and Science, Blekinge Institute
 of Technology,\\ 371 79 Karlskrona, Sweden\\[.5ex]
 {\sc Ranis N. Ibragimov}\\
Department of Mathematics, Research and Support Center for Applied
Mathematical Modeling (RSCAMM),\\
 New Mexico Institute of Mining and Technology,\\
 Socorro, NM, 87801 USA
 \end{center}

 \noindent
\textbf{Abstract.} The paper is devoted to the group
 analysis of equations of motion of two-dimensional uniformly stratified rotating
 fluids used as a basic model in geophysical fluid dynamics. It is shown that
 the nonlinear equations in question have a remarkable property to be self-adjoint.
 This property is crucial for constructing conservation laws provided in the present paper.
 Invariant solutions are constructed using certain symmetries.
 The invariant solutions are used for defining internal wave beams.\\[2ex]

 \section{Introduction}
 \setcounter{equation}{0}


We will apply Lie group analysis for investigating the system of
nonlinear equations
 \begin{align}
 \Delta \psi_t - g \rho_x - f v_z & =
 \psi_x \Delta \psi_z - \psi_z \Delta \psi_x\,, \label{iik.eq1}\\[1ex]
 v_t + f \psi_z & = \psi_x v_z - \psi_z v_x\,, \label{iik.eq2}\\[1ex]
 \rho_t + \frac{N^2}{g}\, \psi_x  & = \psi_x \rho_z - \psi_z \rho_x \label{iik.eq3}
 \end{align}
 used in geophysical fluid dynamics for investigating internal waves in
 uniformly stratified incompressible fluids (oceans). In particular,
 the system (\ref{iik.eq1})-(\ref{iik.eq3}) with $f=0$ was used in
\cite{kis-cha91} to study two non-unidirectional wave beams
propagating and interacting in stratified fluid. An exact solution
of the same system, again in the case when $f = 0,$ was employed
in \cite{lom-ril96} for investigating stability of a singe
internal plane wave. Weakly nonlinear effects in colliding of
 internal wave beams were investigated in \cite{tak03},
 \cite{tal05} by using Eqs. (\ref{iik.eq1})-(\ref{iik.eq3})
 with $f = 0.$ The system (\ref{iik.eq1})-(\ref{iik.eq3})
 with $f \not= 0$ was used in \cite{ibrr10} to model weakly
nonlinear wave interactions governing the time behavior of the
oceanic energy spectrum.

 In these equations $\Delta$ is the
 two-dimensional Laplacian:
 $$
 \Delta = D_x^2 + D_z^2\,, \quad {\rm e.g.} \quad
 \Delta \psi_t = \frac{\partial^2 \psi_t}{\partial x^2} + \frac{\partial^2 \psi_t}{\partial
 z^2} \equiv D_t (\Delta \psi)\,,
 $$
 and  $g, f, N$ are constants. Namely, $g$ is the gravitational
 acceleration, $f$ is the Coriolis parameter. The quantity $N$ appears due
 to the density stratification of a fluid and is constant under
 the linear stratification hypothesis.

 We will  show in what follows that the system of equations (\ref{iik.eq1})-(\ref{iik.eq3})
 is self-adjoint (in the
 terminology of \cite{ibr06a, ibr07a}) and use this remarkable property of the system for calculating
  conservation laws associated with symmetry properties of the system (\ref{iik.eq1})-(\ref{iik.eq3}).

 In some calculations, e.g. in Sections \ref{time-tr}, \ref{flux}, \ref{energy} it is convenient to write Eqs.
 (\ref{iik.eq1})-(\ref{iik.eq3}) by using
  the Jacobians $J(\psi, v) = \psi_x v_z - \psi_z v_x,$ etc.,
   in the following form:
 \begin{align}
 \Delta \psi_t - g \rho_x - f v_z & =
 J(\psi, \Delta \psi), \label{iik.eq1J}\\[1ex]
 v_t + f \psi_z & = J(\psi, v), \label{iik.eq2J}\\[1ex]
 \rho_t + \frac{N^2}{g}\, \psi_x  & = J(\psi, \rho). \label{iik.eq3J}
 \end{align}

 \section{Self-adjointness}
 \label{adj.def}
 \setcounter{equation}{0}

 \subsection{Preliminaries}
  \label{adj:prel}

 We will use the terminology and the following definitions
 from \cite{ibr06a, ibr07a} ~ (see also \cite{ibr06b}).

  Let $x = (x^1, \ldots, x^n)$ be $n$ independent variables, and
$u = (u^1, \ldots, u^m)$ be $m$ dependent variables. The partial
derivatives of $u^{\alpha}$ with respect to $x^i$ are
 denoted by $u_{(1)} = \{u^{\alpha}_i\}, ~ u_{(2)} =
\{u^{\alpha}_{ij}\},\, \ldots$ with
 $$
 u^\alpha_i = D_i(u^\alpha), \quad u^\alpha_{ij} =
D_i(u^\alpha_{j}) = D_iD_j(u^\alpha),\ldots,
 $$
where $D_i$ is the operator of \textit{total differentiation} with
respect to $x^i:$
 \begin{equation}
 \label{aldo1}
  D_i = \frac{\partial}{\partial x^i} +
u^\alpha_{i}\frac{\partial}{\partial u^\alpha} +
u^\alpha_{ij}\frac{\partial}{\partial u^\alpha_j} +
 \cdots, \quad  \ i = 1, \ldots, n.
 \end{equation}
 Even though the operators $D_i$ are given by formal infinite sums, their
 action $D_i(f)$ is well defined for
 functions $f(x, u, u_{(1)}, \ldots)$ depending on
 a finite number of the variables $x, u, u_{(1)}, u_{(2)}, \ldots $
The usual summation convention on repeated indices $\alpha$ and
$i$ is assumed in expressions like Eq. (\ref{aldo1}).

 The variational derivatives (the {\it Euler-Lagrange operator})
are defined by
 \begin{equation}
 \label{el.1}
 \frac{\delta}{\delta u^\alpha} =
\frac{\partial}{\partial u^\alpha} + \sum_{s=1}^\infty (-1)^s
D_{i_1}\cdots D_{i_s}\,\frac{\partial}{\partial
u^\alpha_{i_1\cdots i_s}}\,,\quad \alpha=1, \ldots, m,
\end{equation}
where the summation  over the repeated indices $i_1\ldots i_s$
runs from 1 to $n.$

 \begin{defn}
 \label{adjdef}
 The {\it adjoint equations} to nonlinear partial differential equations
 \begin{equation}
 \label{adjeq1}
 F_\alpha \big(x, u, \ldots, u_{(s)}\big) = 0, \quad
 \alpha = 1, \ldots, m,
 \end{equation}
 are given by  (see also \cite{ath-hom75})
 \begin{equation}
 \label{adjeq2}
 F^*_\alpha \big(x, u, \mu, \ldots, u_{(s)}, \mu_{(s)}\big)  = 0, \quad
 \alpha = 1, \ldots, m,
 \end{equation}
  where $\mu = (\mu^1, \ldots, \mu^m)$ are new dependent variables, and
  $F^*_\alpha$ are defined by
 \begin{equation}
 \label{adjeq0}
 F^*_\alpha \big(x, u, \mu, \ldots, u_{(s)}, \mu_{(s)}\big) =
  \frac{\delta (\mu^\beta F_\beta)}{\delta u^\alpha}\,\cdot
 \end{equation}
 \end{defn}

In the case of linear
 equations, Definition \ref{adjdef} is equivalent to the
 classical definition of the adjoint equation.

 Consider the function
 \begin{equation}
 \label{lagrdef}
 {\cal L} = \mu^\beta F_\beta \big(x, u, \ldots, u_{(s)}\big)
 \end{equation}
 involved in (\ref{adjeq0}). Eqs. (\ref{adjeq1})
 and their adjoint equations (\ref{adjeq2}) can be obtained from (\ref{adjeq0}) by taking the
 variational derivatives (\ref{el.1}) with respect to
 the dependent variables $u$ and the similar variational derivatives
 with respect to the new dependent variables $\mu,$
  \begin{equation}
 \label{el.1coupled}
  \frac{\delta}{\delta \mu^\alpha} =
 \frac{\partial}{\partial \mu^\alpha} + \sum_{s=1}^\infty (-1)^s
 D_{i_1}\cdots D_{i_s}\,\frac{\partial}{\partial
 \mu^\alpha_{i_1\cdots i_s}}\,,\quad \alpha=1, \ldots, m.
 \end{equation}
 Namely:
 \begin{align}
 & \frac{\delta {\cal L}}{\delta \mu^\alpha} = F_\alpha \big(x, u, \ldots,
 u_{(s)}\big),  \label{lagr1} \\[1.5ex]
 &\frac{\delta {\cal L}}{\delta u^\alpha} = F^*_\alpha \big(x, u, \mu, \ldots, u_{(s)},
 \mu_{(s)}\big). \label{lagr2}
 \end{align}
 This circumstance justifies the following definition.

 \begin{defn}
 \label{formlagr}
 The differential function (\ref{lagrdef}) is called a \textit{formal Lagrangian}
 for the differential equations (\ref{adjeq1}).
 For the sake of brevity, formal Lagrangians are also referred to as
 Lagrangians.
  \end{defn}

If the variables $u$ are known, the new variables $\mu$ are
obtained by solving Eqs. (\ref{adjeq2}) which are, according to
(\ref{adjeq0}), linear partial differential equations
(\ref{adjeq2}) with respect to
 $\mu^\alpha.$ Using the existing terminology (see, e.g.
 \cite{akh-gaz89}), we will call $\mu^\alpha$ \textit{nonlocal variables}.

 Nonlocal variables can be excluded from physical quantities such as
  conservation laws if Eqs. (\ref{adjeq1}) are
 \textit{self-adjoint} (\cite{ibr06a}) or, in general,
 \textit{quasi-self-adjoint} (\cite{ibr07d}) in the following
 sense.

 \begin{defn}
 \label{selfadj}
 Eqs. (\ref{adjeq1}) are said to be {\it self-adjoint} if the system obtained from the adjoint equations
 (\ref{adjeq2}) by the substitution $\mu = u:$
 \begin{equation}
 \label{selfadef}
 F^*_\alpha \big(x, u, u, \ldots, u_{(s)}, u_{(s)}\big) = 0, \quad
 \alpha = 1, \ldots, m,
 \end{equation}
 is equivalent to the original system (\ref{adjeq1}), i.e.
 $$
 F^*_\alpha \big(x, u, u, \ldots, u_{(s)}, u_{(s)}\big) =
 \Phi_\alpha^\beta F_\beta \big(x, u, \ldots, u_{(s)}\big),
 \quad \alpha = 1, \ldots, m,
 $$
 with regular (in general, variable) coefficients
 $\Phi_\alpha^\beta.$
 \end{defn}

  \begin{defn}
  \label{q_selfadj}
  Eqs. (\ref{adjeq1}) are said to be quasi-self-adjoint if the
system of adjoint equations  (\ref{adjeq2}) becomes equivalent to
the original system (\ref{adjeq1}) upon the substitution
 \begin{equation}
 \label{qsad:tr}
 \mu = h(u)
 \end{equation}
 with a certain function $h(u)$ such that $h'(u) \not=
0.$
 \end{defn}

 \subsection{Adjoint system to Eqs. (\ref{iik.eq1})-(\ref{iik.eq3})}
 \label{adj}

 Let us apply the methods from Section \ref{adj:prel} to Eqs.
 (\ref{iik.eq1})-(\ref{iik.eq3}). In this case the formal Lagrangian (\ref{lagrdef})
 for Eqs. (\ref{iik.eq1})-(\ref{iik.eq3}) is written
 \begin{equation}
 \label{iik.eq4}
 \begin{aligned}
 {\cal L} & = \varphi \big[\Delta \psi_t - g \rho_x - f v_z -
 \psi_x \Delta \psi_z + \psi_z \Delta \psi_x\big]\\[1.5ex]
 & + \mu \left[v_t + f \psi_z - \psi_x v_z + \psi_z v_x\right]
 + r \Big[\rho_t + \frac{N^2}{g}\, \psi_x - \psi_x \rho_z + \psi_z
 \rho_x\Big],
 \end{aligned}
 \end{equation}
 where $\varphi, \mu$ and $r$ are new dependent variables. The
 adjoint equations to Eqs. (\ref{iik.eq1})-(\ref{iik.eq3}) are
 obtained by taking the variational derivatives of ${\cal L},$
 namely:
 \begin{equation}
 \label{iik.eq5}
 \frac{\delta {\cal L}}{\delta \psi} = 0, \quad
 \frac{\delta {\cal L}}{\delta v} = 0, \quad
 \frac{\delta {\cal L}}{\delta \rho} = 0,
 \end{equation}
 where (see (\ref{el.1}); see also Eqs. (\ref{ii.lem:eq3}))
  \begin{align}
 & \frac{\delta}{\delta v} = \frac{\partial}{\partial v} -
 D_x \frac{\partial}{\partial v_x}
 - D_z \frac{\partial}{\partial v_z}\,,\notag\\[1.5ex]
 & \frac{\delta}{\delta \rho} = \frac{\partial}{\partial \rho} -
 D_x \frac{\partial}{\partial \rho_x} -
 D_z \frac{\partial}{\partial \rho_z}\,,\notag\\[1.5ex]
 & \frac{\delta}{\delta \psi} = \frac{\partial}{\partial \psi} -
 D_x \frac{\partial}{\partial \psi_x}
 - D_z \frac{\partial}{\partial \psi_z}
 + D_x D_t \frac{\partial}{\partial \psi_{xt}}
 + D_z D_t \frac{\partial}{\partial \psi_{zt}}
 + \cdots\,.\notag
 \end{align}

 Taking into account the special form (\ref{iik.eq4}) of ${\cal L},$ we have:
 \begin{align}
 \frac{\delta {\cal L}}{\delta \psi} = & - D_x \frac{\partial {\cal L}}{\partial \psi_x}
 - D_z \frac{\partial {\cal L}}{\partial \psi_z} -
 (D_x^2 + D_z^2) \Big[D_t \frac{\partial {\cal L}}{\partial \Delta \psi_t}
 + D_x \frac{\partial {\cal L}}{\partial \Delta \psi_x} +
  D_z \frac{\partial {\cal L}}{\partial \Delta \psi_z}\Big] \notag\\[1ex]
  = & ~D_x \big(\varphi \Delta \psi_z + \mu v_z - \frac{N^2}{g}\, r + r \rho_z\big)
 - D_z \left(\varphi \Delta \psi_x + f \mu + \mu v_x + r \rho_x\right)\notag\\[1ex]
  & - (D_x^2 + D_z^2) \Big[D_t (\varphi)
 + D_x (\varphi \psi_z) - D_z (\varphi \psi_x)\Big] \notag\\[1ex]
  = & ~\varphi_x \Delta \psi_z  - \varphi_z \Delta \psi_x + \mu_x v_z
   - \frac{N^2}{g}\, r_x + r_x \rho_z
 -  f \mu_z - \mu_z v_x - r_z \rho_x \notag\\[1ex]
  & - \Delta \varphi_t  + 2 \big[\varphi_{xz} \psi_{xx}
 + \varphi_{zz} \psi_{xz} - \varphi_{xx} \psi_{xz}
 - \varphi_{xz} \psi_{zz} \big], \notag\\[1.5ex]
 \frac{\delta {\cal L}}{\delta v} = &
 - D_t \frac{\partial {\cal L}}{\partial v_t} - D_x \frac{\partial {\cal L}}{\partial v_x}
 - D_z \frac{\partial {\cal L}}{\partial v_z} = - \mu_t  - \mu_x \psi_z + f \varphi_z + \mu_z \psi_x\,,\notag\\[1.5ex]
 \frac{\delta {\cal L}}{\delta \rho} = &
 - D_t \frac{\partial {\cal L}}{\partial \rho_t} - D_x \frac{\partial {\cal L}}{\partial \rho_x}
 - D_z \frac{\partial {\cal L}}{\partial \rho_z}= - r_t  + g \varphi_x  - r_x \psi_z + r_z \psi_x\,.\notag
 \end{align}
 Hence, the adjoint equations (\ref{iik.eq5}) can be written as follows:
 \begin{align}
   & \Delta \varphi_t + \frac{N^2}{g}\, r_x +  f \mu_z - \varphi_x \Delta \psi_z
     + \varphi_z \Delta \psi_x - \Theta = 0, \label{iik.eq6} \\[1.5ex]
 & - \mu_t  - \mu_x \psi_z + f \varphi_z + \mu_z \psi_x = 0, \label{iik.eq7}\\[1.5ex]
 & - r_t  + g \varphi_x  - r_x \psi_z + r_z \psi_x = 0, \label{iik.eq8}
 \end{align}
 where
 \begin{equation}
 \label{iik.eq9}
 \Theta = J(\mu, v)
    + J(r, \rho)
    + 2 \big[\varphi_{xz} \psi_{xx}
 + \varphi_{zz} \psi_{xz} - \varphi_{xx} \psi_{xz}
 - \varphi_{xz} \psi_{zz} \big].
 \end{equation}

 \subsection{Self-adjointness of  Eqs. (\ref{iik.eq1})-(\ref{iik.eq3})}

 \begin{thm}
 Eqs. (\ref{iik.eq1})-(\ref{iik.eq3}) are quasi-self-adjoint.
 \end{thm}
 \textbf{Proof.} Looking for (\ref{qsad:tr}) in the form
 of a general scaling transformation, one can readily obtain that after the
 transformation
 \begin{equation}
 \label{iik.eq10}
 \varphi = \psi, \quad \mu = - v, \quad r = - \frac{g^2}{N^2}\,\rho,
 \end{equation}
the quantity $\Theta$ given by Eq. (\ref{iik.eq9}) vanishes.
Therefore the adjoint equations (\ref{iik.eq6})-(\ref{iik.eq8})
become identical with Eqs. (\ref{iik.eq1})-(\ref{iik.eq3}) after
the substitution (\ref{iik.eq10}). Hence, according to Definition
\ref{q_selfadj},  Eqs.
 (\ref{iik.eq1})-(\ref{iik.eq3}) are \textit{quasi-self-adjoint}.
 Since Eqs. (\ref{iik.eq10}) are obtained just be simple scaling
 of the equations $\varphi = \psi,\mu = v, r = \rho$
 required for the self-adjointness, we will say that
   Eqs. (\ref{iik.eq1})-(\ref{iik.eq3}) are \textit{self-adjoint.}

 \section{Conservation laws}
 \label{cl.def}

 \setcounter{equation}{0}

 \subsection{General discussion of conservation equations}

 Along with the individual notation $t, \ x, \ z$ for the the independent
 variables, and $v, \ \rho, \ \psi$ for the dependent variables,
 we will also use the index notation
 $x^1 = t, \ x^2 = x, \ x^3 = z$ and
 $u^1 = v, \ u^2 = \rho, \ u^3 = \psi,$ respectively. We will write the conservation
 laws both in the differential form
 \begin{equation}
 \label{iik.cl_1}
 D_t (C^1) + D_x (C^2) + D_z (C^3) = 0
 \end{equation}
 and the integral form
 \begin{equation}
 \label{iik.cl_2}
 \frac{d}{d t} \int\!\int C^1 dx d z  = 0,
 \end{equation}
 where the double integral in taken over the the $(x, z)$
  plane $\mathbb{R}^2.$
 The equations (\ref{iik.cl_1}) and (\ref{iik.cl_2}) provide a conservation law for
 Eqs. (\ref{iik.eq1})-(\ref{iik.eq3}) if they hold for the
 solutions of Eqs. (\ref{iik.eq1})-(\ref{iik.eq3}). The vector
 $\bm{C} = (C^1, C^2, C^3)$ satisfying the conservation equation (\ref{iik.cl_1})
 is termed a \textit{conserved vector.} Its component $C^1$ is
 called the \textit{density} of the conservation law due to Eq.
 (\ref{iik.cl_2}). The two-dimensional vector $(C^2, C^3)$
 defines  the \textit{flux} of the conservation law.

 The integral form (\ref{iik.cl_2}) of a conservation law
 follows from the differential form (\ref{iik.cl_1})
 provided that the solutions of Eqs. (\ref{iik.eq1})-(\ref{iik.eq3})
  vanish or rapidly decrease at the infinity on
 $\mathbb{R}^2.$ Indeed, integrating Eq. (\ref{iik.cl_1}) over an arbitrary
  region
  $\Omega \subset \mathbb{R}^2$ we have:
 $$
 \frac{d}{d t} \int\!\int_\Omega C^1 dx d z  =
 - \int\!\int_\Omega \left[D_x (C^2) + D_z (C^3)\right] dx d z.
 $$
 According to Green's theorem, the integral on the right-hand side reduces
 to the integral along the boundary $\partial \Omega$ of $\Omega:$
 $$
 - \int\!\int_\Omega \left[D_x (C^2) + D_z (C^3)\right] dx d z
 = \int_{\partial \Omega} C^3 dx - C^2 d z,
 $$
 and hence vanishes as $\Omega$ expands and becomes the plane $\mathbb{R}^2.$

  \begin{rem}
 \label{ii.cl:rem1}
 It is manifest from this discussion that one can ignore in $C^1$
 ``divergent type" terms because they do not change the integral in the
 conservation equation (\ref{iik.cl_2}). Specifically if $C^1$
 evaluated  on the solutions of Eqs.
 (\ref{iik.eq1})-(\ref{iik.eq3}) has the form
 \begin{equation}
 \label{ii.cl_equ}
 C^1 = \widetilde C^1 + D_x(h^2) + D_z(h^3)
 \end{equation}
 with some functions $h^2, h^3,$ then
 the conservation equation (\ref{iik.cl_1}) can be
equivalently rewritten in the form (see \cite{ibr08Selec3}, Paper
1, Section 20.1)
 $$
 D_t (\widetilde C^1) + D_x (\widetilde C^2) + D_z (\widetilde C^3)
 =0,
 $$
 where
 $$
 \widetilde C^2 = C^2 + D_t(h^2), \quad \widetilde C^3 = C^3 + D_t(h^3).
 $$
 Accordingly, we have
 $$
 \int\!\int C^1 dx d z  = \int\!\int \widetilde C^1 dx d z,
 $$
 and hence the integral conservation equation (\ref{iik.cl_2})
 provided by the conservation density $C^1$ of the form
 (\ref{ii.cl_equ}) coincides with that provided by the density $\widetilde C^1.$
 \end{rem}

 In particular, if $\widetilde C^1 = 0$  the integral in Eq. (\ref{iik.cl_2})
 vanishes. This kind of conservation laws are \textit{trivial} from
 physical point of view. Therefore we single out physically useless
 conservation laws by the following definition.

 \begin{defn}
 \label{ii.trcl}
 The conservation law is said to be \textit{trivial}
  if its density $C^1$ evaluated on the solutions
 of Eqs. (\ref{iik.eq1})-(\ref{iik.eq3}) is the divergence,
 $$
 C^1 = D_x(h^2) + D_z(h^3).
 $$
 \end{defn}

 The following statement (\cite{ibr99}, Section 8.4.1; see also \cite{ibr07a})
 simplifies calculations while dealing with conservation
 equations.
 \begin{lem}
 \label{ii.lem}
 A function $F(v, \rho, \psi, v_x, v_z, \rho_x, \rho_z, \psi_x,
 \psi_z, \psi_{xt}, \psi_{zt}, \ldots)$ is the divergence,
 \begin{equation}
 \label{ii.lem:eq1}
 F = D_x(C^1) + D_z(C^2),
 \end{equation}
 if and only if satisfies the following equations:
 \begin{equation}
 \label{ii.lem:eq2}
 \frac{\delta F}{\delta v} = 0, \quad \frac{\delta F}{\delta \rho}
 = 0, \quad
  \frac{\delta F}{\delta \psi} = 0.
 \end{equation}
 Here the variational derivatives act on $F$ as usual (see also Section \ref{adj}):
  \begin{align}
 \label{ii.lem:eq3}
 & \frac{\delta F}{\delta v} = \frac{\partial F}{\partial v} -
 D_x \left(\frac{\partial F}{\partial v_x}\right)
 - D_z \left(\frac{\partial F}{\partial v_z}\right)\,,\notag\\[1.5ex]
 & \frac{\delta F}{\delta \rho} = \frac{\partial F}{\partial \rho} -
 D_x \left(\frac{\partial F}{\partial \rho_x}\right) -
 D_z \left(\frac{\partial F}{\partial \rho_z}\right)\,,\\[1.5ex]
 & \frac{\delta F}{\delta \psi} = \frac{\partial F}{\partial \psi} -
 D_x \left(\frac{\partial F}{\partial \psi_x}\right)
 - D_z \left(\frac{\partial F}{\partial \psi_z}\right)
 + D_x D_t \left(\frac{\partial F}{\partial \psi_{xt}}\right)
 + D_z D_t \left(\frac{\partial F}{\partial \psi_{zt}}\right)
 + \cdots\,.\notag
 \end{align}
 \end{lem}

 \begin{cor}
 \label{ii.lem:cor1}
 A function $C^1$ is the density of a conservation law (\ref{iik.cl_1})
 if and only if the function
 \begin{equation}
 \label{ii.cl_dens}
 F = D_t(C^1)\Big|_{(\ref{iik.eq1})-(\ref{iik.eq3})}
 \end{equation}
 satisfies Eqs. (\ref{ii.lem:eq2}). Here
 $|_{(\ref{iik.eq1})-(\ref{iik.eq3})}$ means that the quantity $D_t(C^1)$ is evaluated on the
 solutions of Eqs. (\ref{iik.eq1})-(\ref{iik.eq3}).
 \end{cor}

 In particular, Lemma \ref{ii.lem} allows one to single out trivial
 conservation laws as follows.
 \begin{cor}
 \label{ii.lem:cor2}
 The conservation law (\ref{iik.cl_1}) is trivial if and only if
 its density $C^1$ evaluated on the solutions of Eqs.
 (\ref{iik.eq1})-(\ref{iik.eq3}), i.e. the quantity
 \begin{equation}
 \label{ii.cl_triv:a}
 C^1_* = C^1\big|_{(\ref{iik.eq1})-(\ref{iik.eq3})}
 \end{equation}
 satisfies Eqs. (\ref{ii.lem:eq2}),
 \begin{equation}
 \label{ii.cl_triv}
 \frac{\delta C^1_*}{\delta v} = 0, \quad
 \frac{\delta C^1_*}{\delta \rho} = 0, \quad
 \frac{\delta C^1_*}{\delta \psi} = 0,
 \end{equation}
 on the solutions of Eqs. (\ref{iik.eq1})-(\ref{iik.eq3}).
 \end{cor}

 \subsection{Variational derivatives of expressions with Jacobians}

 We will use in our calculations the following statement on the behaviour of certain
 expressions with Jacobians under the
 action of the variational derivatives (\ref{ii.lem:eq3}).

 \begin{prop}
 \label{propos}
 The following equations hold:
 \begin{align}
 & \frac{\delta J(\psi, v)}{\delta v} = 0,
 \qquad \frac{\delta J(\psi, v)}{\delta \psi} = 0,\label{ii.vd:1}\\[1.5ex]
 & \frac{\delta [v J(\psi, v)]}{\delta v} = 0,
 \quad \frac{\delta [v J(\psi, v)]}{\delta \psi} = 0,\label{ii.vd:2}\\[1.5ex]
  & \frac{\delta [\rho J(\psi, \rho)]}{\delta \rho} = 0,
 \quad \frac{\delta [\rho J(\psi, \rho)]}{\delta \psi} = 0,\label{ii.vd:3}\\[1.5ex]
 & \frac{\delta J(\psi, \Delta \psi)}{\delta \psi} = 0,
 \quad \frac{\delta [\psi J(\psi, \Delta \psi)]}{\delta \psi} = 0.\label{ii.vd:4}
 \end{align}
 \textbf{Proof.} Let us verify that the first equation (\ref{ii.vd:1}) holds. We have (see (\ref{ii.lem:eq3})):
 $$
 \frac{\delta J(\psi, v)}{\delta v} =
 \frac{\delta (\psi_x v_z - \psi_z v_x)}{\delta v}
 = - D_z(\psi_x) + D_x(\psi_z) = - \psi_{xz} + \psi_{zx} = 0.
 $$
 Replacing $v$ by $\psi$ one obtains the second equation (\ref{ii.vd:1}). Let us
 verify now that Eqs. (\ref{ii.vd:2}) are satisfied. We have:
 \begin{align}
 & \frac{\delta [v J(\psi, v)]}{\delta v} =
 \frac{\delta [v(\psi_x v_z - \psi_z v_x)]}{\delta v}
 = \frac{\partial [v(\psi_x v_z - \psi_z v_x)]}{\partial v} - D_z(v \psi_x) + D_x(v \psi_z)\notag\\[1.5ex]
 & = J(\psi, v) - D_z(v \psi_x) + D_x(v \psi_z)
  = J(\psi, v) - J(\psi, v) - v \psi_{xz} + v \psi{zx} = 0\notag
 \end{align}
 and
 $$
  \frac{\delta [v J(\psi, v)]}{\delta \psi} =
 - D_x(v v_z) + D_z(v v_x) = - v_x v_z - v v_{xz} + v_z v_x + v
 v_{zx} = 0.
 $$
  Replacing $v$ by $\rho$ one obtains Eqs. (\ref{ii.vd:3}).
  Eqs. (\ref{ii.vd:4}) are derived likewise even though they
  involve higher-order derivatives. We have:
 \begin{align}
 \frac{\delta J(\psi, \Delta \psi)}{\delta \psi} & =
 \frac{\delta (\psi_x \Delta \psi_z -
 \psi_z \Delta \psi_x)}{\delta \psi}
 = \frac{\delta [\psi_x (\psi_{xxz} + \psi_{zzz}) -
 \psi_z (\psi_{xxx} + \psi_{xzz})]}{\delta \psi}\notag\\[1.5ex]
 & = - D_x (\Delta \psi_z) + D_z (\Delta \psi_x)
 - D_z (D_x^2 + D_z^2) (\psi_x) + D_x (D_x^2 + D_z^2)(\psi_z)\notag\\[1.5ex]
 & = - D_x (\Delta \psi_z) + D_z (\Delta \psi_x)
 - D_z (\Delta \psi_x) + D_x (\Delta \psi_z) = 0.\notag
 \end{align}
 Derivation of the second equation (\ref{ii.vd:4}) requires only a simple
 modification of the previous calculations. Namely:
  \begin{align}
 & \frac{\delta [\psi J(\psi, \Delta \psi)]}{\delta \psi} =
 \frac{\delta [\psi(\psi_x \Delta \psi_z -
 \psi_z \Delta \psi_x)]}{\delta \psi}\notag\\[1.5ex]
 & = \psi_x \Delta \psi_z - \psi_z \Delta \psi_x
 - D_x (\psi \Delta \psi_z) + D_z (\psi \Delta \psi_x)
 - \Delta D_z (\psi \psi_x) + \Delta D_x (\psi \psi_z)\notag\\[1.5ex]
 & = \psi_x \Delta \psi_z - \psi_z \Delta \psi_x - \psi_x \Delta \psi_z
 - \psi \Delta \psi_{zx} + \psi_z \Delta \psi_x + \psi \Delta \psi_{xz}\notag\\[1.5ex]
 &- \frac{1}{2} \Delta D_z D_x (\psi^2)
 + \frac{1}{2} \Delta D_x D_z (\psi^2) = 0.\notag
 \end{align}
 \end{prop}

 \subsection{Nonlocal conserved vectors}

 It has been demonstrated in \cite{ibr06b, ibr07a}  that for any operator
 \begin{equation}
 \label{ii.cl_sym}
 X = \xi^i \frac{\partial}{\partial x^i} + \eta^\alpha \frac{\partial}{\partial u^\alpha}
 \end{equation}
 admitted by the system (\ref{iik.eq1})-(\ref{iik.eq3}),  the quantities
 \begin{align}
 \label{iik.cleq1}
 &C^i  =\xi^i {\cal L}+W^\alpha\,
 \Big[\frac{\partial {\cal L}}{\partial u_i^\alpha} -
 D_j \Big(\frac{\partial {\cal L}}{\partial u_{ij}^\alpha}\Big)
 + D_j D_k\Big(\frac{\partial {\cal L}}{\partial u_{ijk}^\alpha}\Big)\Big] \qquad \null\\[1.5ex]
 &+D_j\big(W^\alpha\big)\,
 \Big[\frac{\partial {\cal L}}{\partial u_{ij}^\alpha} -
 D_k \Big(\frac{\partial {\cal L}}{\partial
 u_{ijk}^\alpha}\Big)\Big]
 + D_j D_k\big(W^\alpha\big)\Big[\frac{\partial {\cal L}}{\partial u_{ijk}^\alpha}\Big],
 \quad i = 1, 2, 3,\notag
 \end{align}
 define the components of a conserved vector for Eqs.
 (\ref{iik.eq1})-(\ref{iik.eq3}) considered together with the adjoint equations
  (\ref{iik.eq6})-(\ref{iik.eq8}). Here
 \begin{equation}
 \label{iik.cleq1a}
  W^\alpha = \eta^\alpha - \xi^j u_j^\alpha, \quad
  \alpha = 1, 2, 3.
 \end{equation}

 The formula (\ref{iik.cleq1}) is written by taking into account that the Lagrangian
 (\ref{iik.eq4}) involves the derivatives up to third order.
 Moreover, noting that the Lagrangian (\ref{iik.eq4}) vanishes on
 the solutions of Eqs. (\ref{iik.eq1})-(\ref{iik.eq3}), we can
 drop the first term in (\ref{iik.cleq1}) and use the conserved
 vector in the abbreviated form
 \begin{align}
 \label{iik.cleq2}
 &C^i  = W^\alpha\,
 \Big[\frac{\partial {\cal L}}{\partial u_i^\alpha} -
 D_j \Big(\frac{\partial {\cal L}}{\partial u_{ij}^\alpha}\Big)
 + D_j D_k\Big(\frac{\partial {\cal L}}{\partial u_{ijk}^\alpha}\Big)\Big] \qquad \null\\[1.5ex]
 &+D_j\big(W^\alpha\big)\,
 \Big[\frac{\partial {\cal L}}{\partial u_{ij}^\alpha} -
 D_k \Big(\frac{\partial {\cal L}}{\partial
 u_{ijk}^\alpha}\Big)\Big]
 + D_j D_k\big(W^\alpha\big)\,\frac{\partial {\cal L}}{\partial u_{ijk}^\alpha}\,\cdot\notag
 \end{align}

 For computing the conserved vectors (\ref{iik.cleq2}),
  the Lagrangian (\ref{iik.eq4}) containing the mixed derivatives should be written
 in the symmetric form
 \begin{equation}
 \label{iik.Lag}
 \begin{aligned}
 {\cal L} & = \frac{1}{3}\,\varphi \big[\psi_{txx} + \psi_{xtx} + \psi_{xxt}
 + \psi_{tzz} + \psi_{ztz} + \psi_{zzt} - 3 g \rho_x - 3 f v_z\\[1.5ex]
 & - \psi_x \big(\psi_{zxx} + \psi_{xzx} + \psi_{xxz} + 3 \psi_{zzz}\big)
 + \psi_z \big(3 \psi_{xxx} + \psi_{xzz} + \psi_{zxz} + \psi_{zzx}\big)\big]\\[1.5ex]
 & + \mu \left[v_t + f \psi_z - \psi_x v_z + \psi_z v_x\right]
 + r \Big[\rho_t + \frac{N^2}{g}\, \psi_x - \psi_x \rho_z + \psi_z
 \rho_x\Big].
 \end{aligned}
 \end{equation}
 Since the Lagrangian ${\cal L},$ and hence the components (\ref{iik.cleq2}) of a conserved vector
  contain the nonlocal variables $\varphi, \mu, r,$ we obtain in
  this way \textit{nonlocal conserved vectors}.

 \subsection{Computation of nonlocal conserved vectors}

 The substitution of (\ref{iik.Lag}) in
 (\ref{iik.cleq2}) yields:
 \begin{align}
 & C^1 = W^1\, \frac{\partial {\cal L}}{\partial v_t}
 + W^2\, \frac{\partial {\cal L}}{\partial \rho_t} +
  W^3\,\bigg[D_x^2 \bigg(\frac{\partial {\cal L}}{\partial \psi_{txx}}\bigg)
  + D_z^2 \bigg(\frac{\partial {\cal L}}{\partial \psi_{tzz}}\bigg)\bigg]\notag\\[1.5ex]
 & - \bigg[D_x(W^3)\,D_x \bigg(\frac{\partial {\cal L}}{\partial\psi_{txx}}\bigg)
  + D_z(W^3)\, D_z \bigg(\frac{\partial {\cal L}}{\partial \psi_{tzz}}\bigg)\bigg]
  + D_x^2(W^3)\,\frac{\partial {\cal L}}{\partial\psi_{txx}}
  + D_z^2(W^3)\, \frac{\partial {\cal L}}{\partial \psi_{tzz}}\,,\notag
  \end{align}
  or
 \begin{align}
 \label{ii.cl:1}
  C^1 & = W^1\, \mu + W^2\,r +
 \frac{1}{3}\, W^3\,\big[D_x^2 (\varphi) + D_z^2 (\varphi)\big]\\[1.5ex]
 & - \frac{1}{3}\,\Big[\varphi_x\,D_x(W^3) + \varphi_z\,D_z(W^3)\Big]
 + \frac{1}{3}\,\varphi \Big[D_x^2(W^3) + D_z^2(W^3)\Big].\notag
 \end{align}
Furthermore, using the same procedure, we obtain:
 \begin{align}
 C^2 & = W^1\, \frac{\partial {\cal L}}{\partial v_x}
 + W^2\, \frac{\partial {\cal L}}{\partial \rho_x} +
  W^3\,\bigg[\frac{\partial {\cal L}}{\partial \psi_x}
  + D_x^2 \bigg(\frac{\partial {\cal L}}{\partial \psi_{xxx}}\bigg)
  + D_z^2 \bigg(\frac{\partial {\cal L}}{\partial \psi_{xzz}}\bigg)\notag\\[1.5ex]
 & + D_t D_x \bigg(\frac{\partial {\cal L}}{\partial \psi_{xtx}} +
 \frac{\partial {\cal L}}{\partial \psi_{xxt}}\bigg)
 + D_x D_z \bigg(\frac{\partial {\cal L}}{\partial \psi_{xxz}} +
 \frac{\partial {\cal L}}{\partial \psi_{xzx}}\bigg)\bigg]
 - D_x(W^3)\,D_x \bigg(\frac{\partial {\cal L}}{\partial\psi_{xxx}}\bigg)\notag\\[1.5ex]
 & - D_t(W^3)\, D_x \bigg(\frac{\partial {\cal L}}{\partial \psi_{xtx}}\bigg)
  - D_x(W^3)\, D_t \bigg(\frac{\partial {\cal L}}{\partial \psi_{xxt}}\bigg)
 - D_z(W^3)\,D_z \bigg(\frac{\partial {\cal L}}{\partial\psi_{xzz}}\bigg)\notag\\[1.5ex]
 & - D_z(W^3)\, D_x \bigg(\frac{\partial {\cal L}}{\partial \psi_{xzx}}\bigg)
  - D_x(W^3)\, D_z \bigg(\frac{\partial {\cal L}}{\partial \psi_{xxz}}\bigg)
  + D_x^2(W^3)\,\frac{\partial {\cal L}}{\partial\psi_{xxx}}\notag\\[1.5ex]
 & + D_z^2(W^3)\, \frac{\partial {\cal L}}{\partial \psi_{xzz}}
  + D_t D_x (W^3)\bigg(\frac{\partial {\cal L}}{\partial \psi_{xtx}} +
 \frac{\partial {\cal L}}{\partial \psi_{xxt}}\bigg)
 + D_x D_z (W^3) \bigg(\frac{\partial {\cal L}}{\partial \psi_{xxz}} +
 \frac{\partial {\cal L}}{\partial \psi_{xzx}}\bigg),\notag
 \end{align}
 or
 \begin{align}
 \label{ii.cl:2}
 C^2 & = W^1\, \mu \psi_z + W^2\,(r \psi_z - g \varphi) +
  W^3\,\big[- \Delta \psi_z - \mu v_z + \frac{N^2}{g}\,r - r \rho_z\\[1.5ex]
  & + D_x^2 (\varphi \psi_z) + \frac{1}{3}\,D_z^2 (\varphi \psi_z)
   + \frac{2}{3}\,\varphi_{xt}
  - \frac{2}{3}\,D_x D_z(\varphi  \psi_x)\big]\notag\\[1.5ex]
 & - D_x(W^3) \Big[D_x(\varphi \psi_z) + \frac{1}{3}\, \varphi_t - \frac{1}{3}\, D_z(\varphi \psi_x)\Big]
 - \frac{1}{3}\, D_t(W^3)\varphi_x\notag\\[1.5ex]
 & - \frac{1}{3}\, D_z(W^3)\big[D_z(\varphi \psi_z) - D_x(\varphi \psi_x)\big]
 + \Big[D_x^2 (W^3) + \frac{1}{3} D_z^2(W^3)\Big]\varphi \psi_z\notag\\[1.5ex]
 & + \frac{2}{3}\,\varphi D_t D_x (W^3) - \frac{2}{3}\,\varphi \psi_x D_z D_x (W^3).\notag
 \end{align}
 Likewise we get
 \begin{align}
 \label{ii.cl:3}
 C^3 & = - W^1\, (\mu \psi_x + f \varphi) - W^2\,r \psi_x +
  W^3\,\big[- \Delta \psi_x + (f + v_x) \mu +  r \rho_x\\[1.5ex]
  & - \frac{1}{3}\,D_x^2 (\varphi \psi_x) - D_z^2 (\varphi \psi_x)
  + \frac{1}{3}\,D_z^2 (\varphi \psi_z) + \frac{2}{3}\,\varphi_{xt}
  + \frac{2}{3}\,D_x D_z(\varphi  \psi_z)\big]\notag\\[1.5ex]
 & + \frac{1}{3}\, D_x(W^3) \big[D_x(\varphi \psi_x) - D_z(\varphi \psi_z)\big]
 - \frac{1}{3}\, D_t(W^3)\varphi_z\notag\\[1.5ex]
 & - D_z(W^3)\Big[\frac{1}{3}\, \varphi_t - D_z(\varphi \psi_x)
 + \frac{1}{3}\, D_x(\varphi \psi_z)\Big]\notag\\[1.5ex]
 &- \Big[\frac{1}{3} D_x^2 (W^3) + D_z^2(W^3)\Big]\varphi \psi_x
  + \frac{2}{3}\,\varphi D_t D_z (W^3) + \frac{2}{3}\,\varphi \psi_z D_z D_x
 (W^3).\notag
 \end{align}

  \subsection{Local conserved vectors}

 The quantities (\ref{ii.cl:1})-(\ref{ii.cl:3})
 define a nonlocal conserved vector because they contain the nonlocal variables
 $\varphi, \mu, r.$ In consequence, the conservation equation
 (\ref{iik.cl_1}) requires not only the basic equations (\ref{iik.eq1})-(\ref{iik.eq3}),
 but also the adjoint equations (\ref{iik.eq6})-(\ref{iik.eq8}).

 However, we can eliminate the nonlocal variables using the self-adjointness of
 Eqs. (\ref{iik.eq1})-(\ref{iik.eq3}) thus transforming the
 nonlocal conserved vector into a
 local one. Namely, we substitute in Eqs. (\ref{ii.cl:1})-(\ref{ii.cl:3})
 the expressions (\ref{iik.eq10}) for $\varphi, \mu, r:$
 \begin{equation}
 \tag*{(\ref{iik.eq10})}
 \varphi = \psi, \quad \mu = - v, \quad r = - \frac{g^2}{N^2}\,\rho.
 \end{equation}
 Then the adjoint equations (\ref{iik.eq6})-(\ref{iik.eq8}) are
 satisfied for any solutions of the basic equations
 (\ref{iik.eq1})-(\ref{iik.eq3}), and hence the quantities
 (\ref{ii.cl:1})-(\ref{ii.cl:3}) satisfy the conservation equation
 (\ref{iik.cl_1}) on all solutions of Eqs.
 (\ref{iik.eq1})-(\ref{iik.eq3}).

 Let us apply the procedure to
 $C^1.$ We eliminate the nonlocal variables in (\ref{ii.cl:1})
 by substituting there the expressions
 (\ref{iik.eq10}) and write $C^1$ in the following form:
 $$
  C^1 = - v\, W^1 - \frac{g^2}{N^2}\,\rho\, W^2 +
 \frac{1}{3}\big[W^3\,\Delta \psi - \psi_x\,D_x\big(W^3\big) - \psi_z\,D_z\big(W^3\big)
 + \psi \Delta W^3\big],
 $$
 where $$\Delta \psi = D_x^2 (\psi) + D_z^2 (\psi), \quad
  \Delta W^3 = D_x^2 \big(W^3\big) + D_z^2 \big(W^3\big).$$
 We further simplify the expression for $C^1$ by using the identities
 $$
 W^3\,D_x^2 (\psi) = D_x \big[W^3 D_x (\psi)\big] - \psi_x D_x
 \big(W^3\big),
 $$
 $$
 W^3\,D_z^2 (\psi) = D_z \big[W^3 D_z (\psi)\big] - \psi_z D_z \big(W^3\big)
 $$
 and
 $$
 \psi \,D_x^2 \big(W^3\big) = D_x \big[\psi D_x \big(W^3\big)\big] - \psi_x D_x
 \big(W^3\big),
 $$
 $$
 \psi \,D_z^2 \big(W^3\big) = D_z \big[\psi D_z \big(W^3\big)\big] - \psi_z D_z
 \big(W^3\big).
 $$
Then we have:
 \begin{equation}
 \label{ii.cl:1b}
  C^1 = - v\, W^1 - \frac{g^2}{N^2}\,\rho\, W^2 - \psi_x\,D_x\big(W^3\big) - \psi_z\,D_z\big(W^3\big) +
 \frac{1}{3}\,\Delta \big(\psi W^3\big).
 \end{equation}
 Dropping in (\ref{ii.cl:1b}) the divergent type term
 $$
  \frac{1}{3}\,\Delta \big(\psi W^3\big) = D_x \left[\frac{1}{3}\,D_x\big(\psi W^3\big)\right]
 + D_z \left[\frac{1}{3}\,D_z\big(\psi W^3\big)\right]
 $$
 in accordance with Remark \ref{ii.cl:rem1}, we finally obtain:
 \begin{equation}
 \label{ii.cl:1a}
  C^1 = - v\, W^1 - \frac{g^2}{N^2}\,\rho\, W^2 - \psi_x\,D_x\big(W^3\big) - \psi_z\,D_z\big(W^3\big).
 \end{equation}

 We will not dwell on a similar modification of the expressions (\ref{ii.cl:2}),
 (\ref{ii.cl:3}) for the components $C^2$ and $C^3$ of conserved
 vectors. We will see further in Section \ref{flux} that they can be found by simpler
 calculations when a density $C^1$ is known.

 \section{Utilization of obvious symmetries}
 \setcounter{equation}{0}

 \subsection{Introduction}

  Eqs. (\ref{iik.eq1})-(\ref{iik.eq3}) do not
  contain the dependent and independent variables explicitly and therefore
  they are invariant with respect to addition
  of arbitrary constants to all these variables.
 It means that Eqs. (\ref{iik.eq1})-(\ref{iik.eq3}) admit
 the one-parameter groups of translations in all variables,
 $$
 \bar v = v + a_1, \quad \bar \rho = \rho + a_2, \quad \bar \psi = \psi + a_3,
 \quad \bar t = t + a_4, \quad \bar x = x + a_5, \quad \bar z = z + a_6,
 $$
  with the generators
 \begin{equation}
 \label{ii.sym:eq1}
 X_1 = \frac{\partial}{\partial v}\,, \quad
 X_2 = \frac{\partial}{\partial \rho}\,, \quad
 X_3 = \frac{\partial}{\partial \psi}\,, \quad
  X_4 = \frac{\partial}{\partial t}\,, \quad
 X_5 = \frac{\partial}{\partial x}\,, \quad
 X_6 = \frac{\partial}{\partial z}\,\cdot
 \end{equation}
 One can also find by simple calculations the dilations (scaling transformations)
 \begin{equation}
 \label{ii.sym:eq2a}
 \bar v = a v, \quad \bar \rho = b \rho, \quad \bar \psi = c \psi, \quad
 t = \alpha \bar t, \quad x = \beta \bar x, \quad z = \beta \bar z
 \end{equation}
 admitted by  Eqs. (\ref{iik.eq1})-(\ref{iik.eq3}). These transformations are
 defined near the identity transformation if the parameters
 $a, \ldots, \beta$ are positive.
 The dilations of $x$ and $z$ are taken by the same parameter
 $\beta$ in order to keep invariant the Laplacian $\Delta.$
 Let us
 find the parameters $a, \ldots, \beta$ from the invariance condition
 of Eqs. (\ref{iik.eq1})-(\ref{iik.eq3}). The
 transformations (\ref{ii.sym:eq2a}) change the derivatives involved in Eqs.
 (\ref{iik.eq1})-(\ref{iik.eq3}) as follows:
 \begin{align}
 & \bar v_{\bar t} = a \alpha v_t, \quad \bar v_{\bar x} = a \beta v_x, \quad
 \bar v_{\bar z} = a \beta v_z,\notag\\[1ex]
 & \bar \rho_{\bar t} = b \alpha \rho_t, \quad
 \bar \rho_{\bar x} = b \beta \rho_x, \quad \bar \rho_{\bar z} = b \beta
 \rho_z,\label{ii.sym:eq2b}\\[1ex]
 & \bar \psi_{\bar t} = c \alpha \psi_t, \quad \bar \psi_{\bar x} = c \beta \psi_x, \quad
 \bar \psi_{\bar z} = c \beta \psi_z,\notag\\[1ex]
 & \bar \Delta \psi_{\bar t} =
 c \alpha \beta^2 \Delta \psi_t, \ \ \bar \Delta \psi_{\bar x} = c \beta^3 \Delta \psi_x,
 \ \ \bar \Delta \psi_{\bar z} = c \beta^3 \Delta \psi_z,\notag
 \end{align}
where $\bar \Delta$ is the Laplacian written in the variables
$\bar x, \ \bar z.$ The invariance of Eqs.
(\ref{iik.eq1})-(\ref{iik.eq3}) under the transformations
 (\ref{ii.sym:eq2a}) means that the following equations are
 satisfied:
 \begin{align}
 & \bar \Delta \bar \psi_{\bar t} - g \bar \rho_{\bar x} - f \bar v_{\bar z} -
 \bar \psi_{\bar x} \bar \Delta \bar \psi_{\bar z}
 + \bar \psi_{\bar z} \bar \Delta \bar \psi_{\bar x} = 0, \notag\\[1ex]
 & \bar v_{\bar t} + f \bar \psi_{\bar z} - \bar \psi_{\bar x} \bar v_{\bar z}
  + \bar \psi_{\bar z} \bar v_{\bar x} = 0, \notag\\[1ex]
 & \bar \rho_{\bar t} + \frac{N^2}{g}\, \bar \psi_{\bar x}
  - \bar \psi_{\bar x} \bar \rho_{\bar z} + \bar \psi_{\bar z} \bar \rho_{\bar x} = 0, \notag
 \end{align}
 whenever Eqs.(\ref{iik.eq1})-(\ref{iik.eq3}) hold.
 Substituting here the expressions
 (\ref{ii.sym:eq2b}) we have:
  \begin{align}
 & \bar \Delta \bar \psi_{\bar t} - g \bar \rho_{\bar x} - f \bar v_{\bar z} -
 \bar \psi_{\bar x} \bar \Delta \bar \psi_{\bar z}
 + \bar \psi_{\bar z} \bar \Delta \bar \psi_{\bar x} =
 c \alpha \beta^2 \Delta \psi_t - b \beta g \rho_x
 - c^2 \beta^4\big(\psi_x \Delta \psi_z
 - \psi_z \Delta \psi_x\big), \notag\\[1ex]
 & \bar v_{\bar t} + f \bar \psi_{\bar z} - \bar \psi_{\bar x} \bar v_{\bar z}
  + \bar \psi_{\bar z} \bar v_{\bar x} = a \alpha v_t +  c \beta f \psi_z -
  a c \beta^2 \big(\psi_x v_z - \psi_z v_x\big), \notag\\[1ex]
 & \bar \rho_{\bar t} + \frac{N^2}{g}\, \bar \psi_{\bar x}
  - \bar \psi_{\bar x} \bar \rho_{\bar z} + \bar \psi_{\bar z} \bar \rho_{\bar x}
   = b \alpha \rho_t + c \beta \frac{N^2}{g}\, \psi_x
  - b c \beta^2 \big(\psi_x \rho_z - \psi_z \rho_x\big). \notag
 \end{align}
 These equations show that the invariance of Eqs. (\ref{iik.eq1})-(\ref{iik.eq3}) is guaranteed
 by the following six equations for five undetermined parameters $a, b, c, \alpha,
 \beta:$
 \begin{equation}
 \label{ii.sym:eq2c}
  c \alpha \beta^2  = b \beta = c^2 \beta^4, \quad
  a \alpha = c \beta =  a c \beta^2 , \quad
  b \alpha = c \beta = b c \beta^2.
 \end{equation}
 It can be verified by simple calculations that Eqs.
 (\ref{ii.sym:eq2c})
 yield
 $$
 \alpha = 1, \quad  b = a, \quad c = a^2, \quad \beta = \frac{1}{a}\,,
 $$
 where $a$ is an arbitrary parameter. We substitute these values of the parameters
 in (\ref{ii.sym:eq2a}), denote the positive parameter $a$ by ${\rm e}^{\tilde a},$
 drop the tilde and conclude that
  Eqs. (\ref{iik.eq1})-(\ref{iik.eq3})
 admit the one-parameter non-uniform dilation group
 $$
 \bar t = t, \quad \bar x = x {\rm e}^a, \quad \bar z = z {\rm e}^a, \quad
 \bar v = v {\rm e}^a, \quad \bar \rho = \rho {\rm e}^a, \quad
 \bar \psi = \psi {\rm e}^{2a}
 $$
 with the following generator:
 \begin{equation}
 \label{ii.sym:eq2}
 X_7 = x \frac{\partial}{\partial x} + z \frac{\partial}{\partial z}
 + v \frac{\partial}{\partial v} + \rho \frac{\partial}{\partial \rho}
 + 2 \psi \frac{\partial}{\partial \psi}\,\cdot
 \end{equation}
 We will consider the operators (\ref{ii.sym:eq1})-(\ref{ii.sym:eq2}) as obvious
 symmetries of Eqs. (\ref{iik.eq1})-(\ref{iik.eq3}) and will
 compute the local conservation laws provided by these symmetries.

 \subsection{Translation of $\bm{v}$}

 For the operator $X_1$ from (\ref{ii.sym:eq1})  Eqs.
 (\ref{iik.cleq1a}) yield
 $$
 W^1 = 1, \quad W^2 = 0, \quad W^3 = 0.
 $$
 Substituting these expressions in Eq. (\ref{ii.cl:1a}) we obtain
 $$
 C^1 = - v.
 $$
 In this case the equations (\ref{ii.cl:2}) and (\ref{ii.cl:3}) are also
 simple. They are written
 $$
 C^2 = u \psi_z, \quad
 C^3 = - (u \psi_x + f \varphi)
 $$
 and upon using Eqs. (\ref{iik.eq10}) yield:
 $$
 C^2 = - v \psi_z, \quad
 C^3 = v \psi_x - f \psi.
 $$
 Since any conserved vector is defined up to multiplication by an
 arbitrary constant, we change the sign of $C^1, \ C^2, \ C^3$ and
 obtain the following conserved vector:
 \begin{equation}
 \label{ii.sym:eq3}
 C^1 = v, \quad C^2 = v \psi_z, \quad
 C^3 = f \psi - v \psi_x.
 \end{equation}
 We have:
 $$
 D_t (C^1) + D_x (C^2) + D_z (C^3) = v_t + v_x \psi_z + f \psi_z - v_z \psi_x.
 $$
 Hence, the conservation equation (\ref{iik.cl_1}) coincides with Eq. (\ref{iik.eq2}).

 \subsection{Translation of $\bm{\rho}$}

  For the operator $X_2$ from (\ref{ii.sym:eq1})  Eqs.
 (\ref{iik.cleq1a}) yield
 $$
 W^1 = 0, \quad W^2 = 1, \quad W^3 = 0.
 $$
 Substituting these expressions in Eq. (\ref{ii.cl:1a}) we obtain
 $$
 C^1 = - \frac{g^2}{N^2}\,\rho.
 $$
 Furthermore, Eqs. (\ref{ii.cl:2}), (\ref{ii.cl:3}) and Eqs. (\ref{iik.eq10}) yield:
 $$
 C^2 = - g \psi - \frac{g^2}{N^2}\, \rho\, \psi_z, \quad
 C^3 = \frac{g^2}{N^2}\, \rho\, \psi_x.
 $$
 Multiplying $C^1, \ C^2, \ C^3$ by $- N^2/ g^2$ we arrive at
 the following conserved vector:
 \begin{equation}
 \label{ii.sym:eq4}
 C^1 = \rho, \quad C^2 = \frac{N^2}{g}\, \psi + \rho\, \psi_z, \quad
 C^3 =  - \rho\, \psi_x.
 \end{equation}
 One can readily verify that the conservation equation (\ref{iik.cl_1})
 for the vector (\ref{ii.sym:eq4}) is also satisfied. Namely, it
 coincides with Eq. (\ref{iik.eq3}).

  \subsection{Translation of $\bm{\psi}$}

  For the operator $X_3$ from (\ref{ii.sym:eq1})  Eqs.
 (\ref{iik.cleq1a}) yield
 $$
 W^1 = 0, \quad W^2 = 0, \quad W^3 = 1.
 $$
 Substituting these expressions in Eq. (\ref{ii.cl:1a}) we obtain
 $$
 C^1 = 0.
 $$
 Hence, the invariance of Eqs. (\ref{iik.eq1})-(\ref{iik.eq3})
 under the translation of $\psi$ furnishes only a trivial
 conservation law (see Definition \ref{ii.trcl}).

 \subsection{Derivation of the flux of conserved vectors with known densities}
 \label{flux}

 We will show here how to find the components $C^2$ and $C^3$ of
 the conserved vector (\ref{ii.sym:eq3}) without using
 Eqs. (\ref{ii.cl:2}), (\ref{ii.cl:3}), provided
 that we know the conserved density $C^1 = v.$

  Let us first verify that $C^1 = v$ satisfies Corollary
 \ref{ii.lem:cor1}. In this case $D_t(C^1) = v_t,$  and hence Eq. (\ref{ii.cl_dens})
 yields
 \begin{equation}
 \label{ii.sym:eq3a}
 F = D_t(C^1)\Big|_{(\ref{iik.eq1})-(\ref{iik.eq3})} =
  J(\psi, v) - f \psi_z.
 \end{equation}
 Using Proposition \ref{propos} we see that Eqs. (\ref{ii.lem:eq2}) are satisfied:
 \begin{equation}
  \frac{\delta F}{\delta v} = \frac{\delta F}{\delta \rho}
  = 0,\quad
  \frac{\delta F}{\delta \psi} = D_z(f) = 0.
 \end{equation}
 Therefore Corollary \ref{ii.lem:cor1} guarantees that $F$ defined by Eq.
 (\ref{ii.sym:eq3a}) satisfies Eq. (\ref{ii.lem:eq1}):
 \begin{equation}
 \label{ii.sym:eq3b}
 \psi_x v_z - \psi_z v_x - f \psi_z = D_x(H^1) + D_z(H^2)
 \end{equation}
 with certain functions $H^1, \ H^2.$

 In order to find $H^1, \ H^2,$ we write
 $$
 \psi_x v_z - f \psi_z = D_z(v \psi_x - f \psi) - v \psi_{xz}, \quad
  - \psi_z v_x = D_x(- v \psi_z) + v \psi_{zx}
 $$
 and obtain:
 $$
 \psi_x v_z - \psi_z v_x - f \psi_z = D_x(- v \psi_z) +
 D_z(v \psi_x - f \psi).
 $$
 Thus, $H^1 = - v \psi_z, \ H^2 = v \psi_x - f \psi.$ Denoting
 $C^2 = - H^1, \ C^3 = - H^2,$ i.e.
 $$
 C^1 =  v \psi_z, \quad  C^2 = f \psi - v \psi_x,
 $$
 and invoking Eq. (\ref{ii.sym:eq3a}), we write Eq. (\ref{ii.sym:eq3b}) in
 the form
 $$
 D_t(C^1)\Big|_{(\ref{iik.eq1})-(\ref{iik.eq3})} + D_x(C^1) +
 D_z(C^2)=0.
 $$
 This is precisely the conservation equation (\ref{iik.cl_1})
 for the vector (\ref{ii.sym:eq3}).
 Thus, we have obtained the components $C^2, \ C^3$ of the
 conserved vector (\ref{ii.sym:eq3}) without using Eqs. (\ref{ii.cl:2}),
 (\ref{ii.cl:3}).

 The components $C^2, \ C^3$ of the
 conserved vector (\ref{ii.sym:eq4}) can be derived likewise.

 \subsection{Translation of $\bm{x}$}

 For the operator $X_5$ from (\ref{ii.sym:eq1})  Eqs.
 (\ref{iik.cleq1a}) yield
 $$
 W^1 = - v_x, \quad W^2 = - \rho_x, \quad W^3 = - \psi_x.
 $$
 Substituting these expressions in Eq. (\ref{ii.cl:1a}) we obtain
 $$
 C^1 = v v_x + \frac{g^2}{N^2}\, \rho \rho_x + \psi_x \psi_{xx} + \psi_z \psi_{xz}
 = D_x\left(\frac{1}{2} v^2 + \frac{1}{2} \frac{g^2}{N^2}\, \rho^2
  + \frac{1}{2} \psi_x^2 + \frac{1}{2} \psi_z^2\right).
 $$
 Hence, the invariance of Eqs. (\ref{iik.eq1})-(\ref{iik.eq3})
 under the translation of $x$ furnishes only a trivial
 conservation law (see Definition \ref{ii.trcl}). Similar
 calculations show that the invariance under the translation of $z$
 provides also a trivial conservation law.

  \subsection{Time translation}
  \label{time-tr}

 For the operator $X_4$ from (\ref{ii.sym:eq1})  Eqs.
 (\ref{iik.cleq1a}) yield
 $$
 W^1 = - v_t, \quad W^2 = - \rho_t, \quad W^3 = - \psi_t.
 $$
 Substituting these expressions in Eq. (\ref{ii.cl:1a}) we obtain
 \begin{equation}
 \label{ii.time.den}
 C^1 = v v_t + \frac{g^2}{N^2}\, \rho \rho_t + \psi_x \psi_{xt} + \psi_z \psi_{zt}.
 \end{equation}
 Changing the last two terms of $C^1$ by using the identity
 \begin{align}
 \label{ii.time.iden}
 \psi_x \psi_{xt} + \psi_z \psi_{zt} & = D_x(\psi \psi_{xt}) - \psi \psi_{xxt}
 + D_z(\psi \psi_{zt}) - \psi \psi_{zzt}\notag\\
 & = D_x(\psi \psi_{xt}) + D_z(\psi \psi_{zt}) - \psi \Delta \psi_t
 \end{align}
 and dropping the divergent type terms, we rewrite $C^1$ given by Eq. (\ref{ii.time.den}) in the
 form
 \begin{equation}
 \label{ii.sym:eq3c}
 C^1 = v v_t + \frac{g^2}{N^2}\, \rho \rho_t - \psi \Delta \psi_t.
 \end{equation}

 Let us clarify if the conservation
 law with the density (\ref{ii.sym:eq3c}) is trivial or non-trivial.
 According to Definition \ref{ii.trcl}, we have to evaluate the density
 (\ref{ii.sym:eq3c})
 on the solutions of Eqs. (\ref{iik.eq1})-(\ref{iik.eq3}).
 In this case it is convenient to use Eqs. (\ref{iik.eq1})-(\ref{iik.eq3}) in the form
 (\ref{iik.eq1J})-(\ref{iik.eq3J}) and replace
  Eq. (\ref{ii.cl_triv:a}) by
  $$
  C^1_* = C^1\big|_{(\ref{iik.eq1J})-(\ref{iik.eq3J})}.
  $$
 Then we have
 $$
 C^1_* = \Big\{v J(\psi, v) + \frac{g^2}{N^2}\,\rho J(\psi, \rho)
 - \psi J(\psi, \Delta \psi)\Big\}- f D_z(v \psi) - g D_x(\rho \psi)
 $$
 and Corollary \ref{ii.lem:cor2} shows that the conservation law
 is trivial. Indeed, the last two
 terms of $C^1_*$ have the divergent form. The expression in
 braces for $C^1_*$ satisfies Eqs.
 (\ref{ii.cl_triv}) according to Proposition \ref{propos}, and hence
it also has the divergent form.

 Thus, the invariance of Eqs. (\ref{iik.eq1})-(\ref{iik.eq3})
 under the time translation furnishes only a trivial
 conservation law.

 \subsection{Use of the dilation. Conservation of energy}
 \label{energy}

 Consider the generator (\ref{ii.sym:eq2}) of the dilation group,
 $$
 X_ 7 = x \frac{\partial}{\partial x} + z \frac{\partial}{\partial z}
 + v \frac{\partial}{\partial v} + \rho \frac{\partial}{\partial \rho}
 + 2 \psi \frac{\partial}{\partial \psi}\,\cdot
 $$
 In this case the quantities (\ref{iik.cleq1a}) have the form
 \begin{equation}
 \label{iik.cleq1b}
 W^1 = v - x v_x - z v_z, \quad W^2 = \rho - x \rho_x - z \rho_z,
  \quad W^3 = 2 \psi - x \psi_x - z \psi_z.
 \end{equation}
 The substitution of (\ref{iik.cleq1b}) in (\ref{ii.cl:1a})
 yields:
 \begin{align}
 \label{ii.dil:1}
 &C^1  = - v^2 + x v v_x + z v v_z
 + \frac{g^2}{N^2}\left(- \rho^2 + x \rho \rho_x + z \rho \rho_z\right)\notag\\[1.5ex]
 & - \psi_x^2 + x \psi_x \psi_{xx} + z \psi_x \psi_{xz}
 - \psi_z^2 + x \psi_z \psi_{xz} + z \psi_z \psi_{zz}.
 \end{align}
 We modify (\ref{ii.dil:1}) by using the identities
 \begin{align}
 & x v v_x + z v v_z = \frac{1}{2}\,D_x\left(x v^2\right)
 + \frac{1}{2}\,D_z\left(z v^2\right) - v^2,\notag\\[1ex]
 & x \rho \rho_x + z \rho \rho_z = \frac{1}{2}\,D_x\left(x \rho^2\right)
 + \frac{1}{2}\,D_z\left(z \rho^2\right) - \rho^2,\notag\\[1ex]
 & x \psi_x \psi_{xx} + x \psi_z \psi_{xz} =
 \frac{1}{2}\,D_x\left[x\left(\psi_x^2 + \psi_x^2\right)\right]
 - \frac{1}{2}\,\left(\psi_x^2 + \psi_x^2\right),\notag\\[1ex]
 & z \psi_x \psi_{xz} + z \psi_z \psi_{zz} =
 \frac{1}{2}\,D_z\left[z\left(\psi_x^2 + \psi_x^2\right)\right]
 - \frac{1}{2}\,\left(\psi_x^2 + \psi_x^2\right).\notag
 \end{align}
 Substituting these in (\ref{ii.dil:1}) and dropping the divergent type terms we have:
 $$
  C^1 = - 2\left(v^2 + \frac{g^2}{N^2}\,\rho^2 + |\nabla \psi|^2\right),
 $$
  where $$|\nabla \psi|^2 = \psi_x^2 + \psi_z^2.$$
 Dividing $C^1$ by the inessential coefficient $(- 2)$ we finally obtain
 the following conservation law in the integral form (\ref{iik.cl_2}):
 \begin{equation}
 \label{iik.cl_dil:eq2}
 \frac{d}{d t} \int\!\int \left[v^2 + \frac{g^2}{N^2}\, \rho^2
 + |\nabla \psi|^2\right] dx d z  = 0.
 \end{equation}
 Eq. (\ref{iik.cl_dil:eq2}) represents the conservation of the \textit{energy with the density}
 \begin{equation}
 \label{ii.ener-dens}
 E = v^2 + \frac{g^2}{N^2}\,\rho^2 + |\nabla \psi|^2.
 \end{equation}

 Let us find the components $C^2$ and $C^3$ of this conservation law written in the differential
  form (\ref{iik.cl_1}). We will use the procedure suggested in
  Section \ref{flux}. Let us first verify that $E$ defined by Eq. (\ref{ii.ener-dens})
  satisfies Corollary \ref{ii.lem:cor1} for densities of conservation laws.
 We have
 \begin{equation}
 \label{ii.ener-dens:eq0}
 D_t(E) = 2 \left(v v_t + \frac{g^2}{N^2}\,\rho \rho_t + \psi_x \psi_{xt} + \psi_z
 \psi_{zt}\right).
 \end{equation}
 Since the expression in the brackets in Eq.
 (\ref{ii.ener-dens:eq0}) is identical with
 (\ref{ii.time.den}) it can be rewritten in the form
 (\ref{ii.sym:eq3c}), and hence satisfies Eqs. (\ref{ii.lem:eq2}).
  Corollary \ref{ii.lem:cor1} guarantees that $E$ is the density of a conservation
  law. It is manifest from Eq. (\ref{ii.ener-dens}) that this conservation law is  non-trivial.

 According to Corollary \ref{ii.lem:cor1}, $D_t(E)$ defined by Eq.
 (\ref{ii.ener-dens:eq0}) and evaluated on
 the solutions of Eqs. (\ref{iik.eq1})-(\ref{iik.eq3}) satisfies Eq. (\ref{ii.lem:eq1}),
 \begin{equation}
 \label{ii.ener-dens:eq1}
 D_t(E)\Big|_{(\ref{iik.eq1})-(\ref{iik.eq3})} =
 D_x(H^1) + D_z(H^2),
 \end{equation}
 with certain functions $H^1, \ H^2.$
 In order to find $H^1, \ H^2,$ we use Eq. (\ref{ii.time.iden}),
 \begin{equation}
 \label{ii.time.iden:a}
 2(\psi_x \psi_{xt} + \psi_z \psi_{zt}) = D_x(2 \psi \psi_{xt})
 + D_z(2 \psi \psi_{zt}) - 2 \psi \Delta \psi_t,
 \end{equation}
 and write:
 \begin{align}
 \label{ii.ener-dens:eq2}
 2 v v_t & = 2 \psi_x v v_z - \psi_z v v_x - 2 f v \psi_z\notag\\
 & = D_z (v^2 \psi_x) - D_x (v^2 \psi_z)
   - 2 f D_z (v \psi) + 2 f \psi v_z,
 \end{align}
  \begin{align}
 \label{ii.ener-dens:eq3}
 2\,\frac{g^2}{N^2}\, \rho \rho_t & = 2\,\frac{g^2}{N^2}\,
 \Big(\psi_x \rho \rho_z - \psi_z \rho \rho_x\Big) - 2 g \rho \psi_x\notag\\
 & = \frac{g^2}{N^2}\,\Big[D_z (\rho^2 \psi_x) - D_x (\rho^2
 \psi_z)\Big] - 2 g D_x (\rho \psi) + 2 g \psi \rho_x,
 \end{align}
 \begin{align}
 \label{ii.ener-dens:eq4}
 - 2 \psi \Delta \psi_t & =
  - 2 g \psi \rho_x - 2 f \psi v_z - 2 \psi \psi_x \Delta \psi_z
  + 2 \psi \psi_z \Delta \psi_x\notag\\
 & = - 2 g \psi \rho_x - 2 f \psi v_z - D_x\big(\psi^2 \Delta \psi_z\big)
  + D_z\big(\psi^2 \Delta \psi_x\big).
 \end{align}
 Substituting the expressions (\ref{ii.ener-dens:eq2}), (\ref{ii.ener-dens:eq3})
 and (\ref{ii.time.iden:a}), (\ref{ii.ener-dens:eq4}) in the right-hand side
 of Eq. (\ref{ii.ener-dens:eq0}), we arrive at Eq.
 (\ref{ii.ener-dens:eq1}) with
 \begin{align}
 & H^1 = - v^2 \psi_z - \frac{g^2}{N^2}\,\rho^2 \psi_z
 - 2 g \rho \psi + 2 \psi \psi_{xt} - \psi^2 \Delta \psi_z,\notag\\
 & H^2 = v^2 \psi_x + \frac{g^2}{N^2}\,\rho^2 \psi_x
 - 2 f v \psi + 2 \psi \psi_{zt} + \psi^2 \Delta \psi_x.\notag
 \end{align}
 Thus, denoting $C^2 = - H^1, \ C^3 = - H^2$ we arrive at the
 following differential form (\ref{iik.cl_1}) of the conservation of energy
 for Eqs. (\ref{iik.eq1})-(\ref{iik.eq3}):
 \begin{equation}
 \label{ii.ener-cons:eq1}
 D_t(E) + D_x(C^2) + D_x(C^3) = 0
 \end{equation}
 with the density $E$ given by Eq. (\ref{ii.ener-dens}) and the
 flux given by the equations
 \begin{align}
  \label{ii.ener-cons:eq2}
 & C^2 = 2 g \rho \psi + v^2 \psi_z + \frac{g^2}{N^2}\,\rho^2 \psi_z
  - 2 \psi \psi_{xt} + \psi^2 \Delta \psi_z,\notag\\
 & C^3 = 2 f v \psi - v^2 \psi_x - \frac{g^2}{N^2}\,\rho^2 \psi_x
  - 2 \psi \psi_{zt} - \psi^2 \Delta \psi_x.
 \end{align}

 \section{Invariant solutions}
 \setcounter{equation}{0}

 \subsection{Invariant solution based on translation and dilation}

 Let us find the invariant solution based on the following two
 operators:
 \begin{equation}
 \label{ig.eq6}
 \begin{split}
 & m X_5 - k X_6 = m \frac{\partial}{\partial x}
 - k \frac{\partial}{\partial z} \quad
 (m, k = {\rm const.}),\\[1ex]
 & X_7 = x \frac{\partial}{\partial x} + z \frac{\partial}{\partial z}
 + v \frac{\partial}{\partial v} + \rho \frac{\partial}{\partial \rho}
 + 2 \psi \frac{\partial}{\partial \psi}\,\cdot
 \end{split}
 \end{equation}
 We first find their invariants $J(t, x, z, v, \rho, \psi)$ by
 solving the equations
 \begin{equation}
 \label{ig.eq6:inv1}
 (m X_5 - k X_6) J = 0, \quad X_7 J = 0.
 \end{equation}
 The characteristic equation $k dx + m dz = 0$ for the first equation (\ref{ig.eq6:inv1})
 yields that the operator $m X_2 - k X_3$ has, along with $t, v, \rho,
 \psi,$ the following invariant:
 \begin{equation}
 \label{ig.eq6:inv2}
 \lambda = k x + m z.
 \end{equation}
 Therefore we have to find the invariants $J(t, \lambda, v, \rho, \psi)$
 for the operator $X_7.$ To this end, we write the action of
 $X_7$ on the variables $t, \lambda, v, \rho, \psi$ by the
 standard formula
 $$
 X_7 = X_7(\lambda) \frac{\partial}{\partial \lambda}
 + v \frac{\partial}{\partial v} + \rho \frac{\partial}{\partial \rho}
 + 2 \psi \frac{\partial}{\partial \psi}
 $$
and obtain
 \begin{equation}
 \label{ig.eq6:inv3}
 X_7 = \lambda \frac{\partial}{\partial \lambda}
 + v \frac{\partial}{\partial v} + \rho \frac{\partial}{\partial \rho}
 + 2 \psi \frac{\partial}{\partial \psi}\,\cdot
 \end{equation}

 To solve the equation $X_7 J(t, \lambda, v, \rho, \psi) = 0$ for the invariants,
 we calculate the first integrals for the characteristic system
 $$
 \frac{d \lambda}{\lambda} = \frac{d v}{v} = \frac{d \rho}{\rho} =
 \frac{d \psi}{2 \psi}
 $$
 and see that a basis of invariants for the operators
 (\ref{ig.eq6}) is given by
 $$
 t, \quad V = \frac{v}{\lambda}\,, \quad R = \frac{\rho}{\lambda}\,, \quad
 \phi = \frac{\psi}{\lambda^2}\,\cdot
 $$
 Accordingly, we assign the invariants $V, R, \phi$ to be functions of the
 invariant $t$ and arrive at the following general form of the
 candidates for the invariant solutions:
 \begin{equation}
 \label{ig.eq7}
 v = \lambda V(t), \quad \rho = \lambda R(t), \quad
 \psi = \lambda^2 \phi(t), \quad \lambda = k x + m z.
 \end{equation}
  In order to find the functions $V(t), R(t), \phi(t),$
  we have to substitute the expressions (\ref{ig.eq7}) in Eqs. (\ref{iik.eq1})-
 (\ref{iik.eq3}).

 We have:
 \begin{align}
 & \psi_t = \lambda^2 \phi'(t), \quad \psi_x = 2 k \lambda \phi(t),
 \quad \psi_z = 2 m \lambda \phi(t),\notag\\[1ex]
 & \nabla^2 \psi_t = 2(k^2 + m^2) \phi'(t), \quad \nabla^2 \psi_x = 0,
 \quad \nabla^2 \psi_z = 0,\notag\\[1ex]
 & \psi_x v_z = 2 k m \lambda\, \phi(t)\, V(t),
 \quad \psi_z v_x = 2 k m \lambda\, \phi(t)\, V(t),\notag\\[1ex]
  & \psi_x \rho_z = 2 k m \lambda\, \phi(t)\, R(t),
 \quad \psi_z \rho_x = 2 k m \lambda\, \phi(t)\, R(t).\notag
 \end{align}
 Therefore Eqs. (\ref{iik.eq1})- (\ref{iik.eq3}) yield the following
 system of first-order linear ordinary differential equations:
 \begin{align}
 2 \big(k^2 + m^2\big)\phi' - g k R - f m V & = 0, \notag\\[1ex]
 \lambda V' + 2 f m \lambda \phi & = 0, \notag\\[1ex]
 \lambda R' + 2 \frac{k \lambda}{g}\,  N^2 \phi & = 0,
 \notag
 \end{align}
 or
 \begin{align}
 \phi' & = \frac{1}{ 2 \big(k^2 + m^2\big)} \big(g k R + f m V\big), \label{ig.eq8}\\[1ex]
 V' & = - 2f m \,\phi, \label{ig.eq9}\\[1ex]
 R' & = - 2 \frac{k}{g}\, N^2 \phi\,. \label{ig.eq10}
 \end{align}

 Let us integrate Eqs. (\ref{ig.eq8})-(\ref{ig.eq10}).
 Differentiating Eq. (\ref{ig.eq8}) and using Eqs.
 (\ref{ig.eq9})-(\ref{ig.eq10}), we obtain
 \begin{equation}
 \label{ig.eq11}
 \phi'' + \omega^2\, \phi = 0,
 \end{equation}
 where
 \begin{equation}
 \label{ig.eq13}
 \omega^2 = \frac{k^2 N^2 + m^2f^2}{k^2 + m^2}\,.
 \end{equation}

 The general solution of Eq. (\ref{ig.eq11}) is given by
 \begin{equation}
 \label{ig.eq12}
 \phi(t) = C_1 \cos (\omega t) + C_2 \sin (\omega t), \quad C_1, C_2 = {\rm
 const.}
 \end{equation}
Substituting (\ref{ig.eq12}) in Eqs.
(\ref{ig.eq9})-(\ref{ig.eq10}) and integrating, we obtain
 \begin{align}
  V & = C_3 - \frac{2 f m}{\omega} \Big[C_1 \sin (\omega t) - C_2 \cos (\omega t)\Big], \notag\\[1ex]
   R & = C_4 - \frac{2k}{g \omega}\, N^2 \Big[C_1 \sin (\omega t) - C_2 \cos (\omega t)\Big].\notag
 \end{align}
 To determine the constants $C_3$ and $C_4,$ we substitute in Eq.
(\ref{ig.eq8}) the above expressions for $V, R$ and the expression
(\ref{ig.eq12}) for $\phi$ and obtain
 $$
 f m C_3 + g k C_4 = 0.
 $$
 Thus,
the solution to Eqs. (\ref{ig.eq8})-(\ref{ig.eq10}) has the
following form:
 \begin{align}
 \phi(t) & = C_1 \cos (\omega t) + C_2 \sin (\omega t), \label{ig.eq14}\\[1ex]
 V (t) & = \frac{2 f m}{\omega} \Big[C_2 \cos (\omega t) - C_1 \sin (\omega t)\Big] + C_3, \label{ig.eq15}\\[1ex]
 R (t) & = \frac{2 k}{g\, \omega}\, N^2 \Big[C_2 \cos (\omega t) - C_1 \sin (\omega t)\Big]
 - \frac{f m}{g k}\,C_3. \label{ig.eq16}
 \end{align}

 Finally, substituting (\ref{ig.eq14})-(\ref{ig.eq16}) in
 (\ref{ig.eq7}), we arrive at the following solution to the system
 (\ref{iik.eq1})-(\ref{iik.eq3}):
 \begin{align}
 \rho & = \frac{2 k}{g\, \omega}\, N^2 \Big[C_2 \cos (\omega t)
 - C_1 \sin (\omega t)\Big] \lambda - \frac{f m}{g k}\,C_3 \lambda, \label{ig.eq17}\\[1ex]
 v & = \frac{2 f m}{\omega} \Big[C_2 \cos (\omega t)
 - C_1 \sin (\omega t)\Big] \lambda + C_3 \lambda, \label{ig.eq18}\\[1ex]
 \psi & = \big[C_1 \cos (\omega t) + C_2 \sin (\omega t)\big] \lambda^2, \label{ig.eq19}
 \end{align}
 where $\lambda$ is given by (\ref{ig.eq6:inv2}), $\omega$ is defined by Eq. (\ref{ig.eq13}) and $C_1, C_2,
 C_3$ are arbitrary constants.

 \subsection{Generalized invariant solution and wave beams}

 It is natural to generalize the candidates (\ref{ig.eq7}) for the invariant
 solutions and look for particular solutions of the system (\ref{iik.eq1})-(\ref{iik.eq3})
 in the following form of separated variables:
 \begin{equation}
 \label{ig.eq20}
 v = F(\lambda) V(t), \quad \rho = \alpha(\lambda) R(t), \quad
 \psi = \beta(\lambda) \phi(t), \quad \lambda = k x + m z.
 \end{equation}

 The reckoning shows that then the right-hand sides of Eqs. (\ref{iik.eq1})-(\ref{iik.eq3})
 vanish and Eqs. (\ref{iik.eq1})-(\ref{iik.eq3}) become:
 \begin{align}
 \big(k^2 + m^2\big)\beta''(\lambda) \phi'(t) - g k \alpha\,'(\lambda)  R(t)
 - f m F'(\lambda) V(t) & = 0, \label{ig.eq21}\\[1ex]
 F (\lambda) V'(t) + f m \beta'(\lambda) \phi(t) & = 0, \label{ig.eq22}\\[1ex]
 \alpha(\lambda)R'(t) + \frac{k N^2}{g}\, \beta'(\lambda) \phi(t) & = 0. \label{ig.eq23}
 \end{align}
 Differentiating Eq. (\ref{ig.eq21}) with respect to $t,$ using
 Eqs. (\ref{ig.eq22})-(\ref{ig.eq23}) and dividing by $\beta',$ we obtain
 $$
 \big(k^2 + m^2\big) \frac{\beta''}{\beta'}\, \phi''
 + \left(N^2 k^2 \frac{\alpha\,'}{\alpha}
 + f^2 m^2 \frac{F'}{F}\right) \phi = 0.
 $$
 Assuming that the ratios $\beta''/\beta'\,, \ \alpha\,'/\alpha, \
 F'/F$ are proportional with constant coefficients and have one
 and the same sign, we arrive at an equation of the form (\ref{ig.eq11}). For example, letting
 \begin{equation}
 \label{ig.eq24}
 \frac{\beta''}{\beta'} = \frac{\alpha\,'}{\alpha}
 =\frac{F'}{F}\,,
 \end{equation}
 we obtain Eq. (\ref{ig.eq11}). Then, according to
 (\ref{ig.eq12}), we can set in (\ref{ig.eq20}) $\phi (t) = \cos (\omega t)$
 and $\phi (t) = \sin (\omega t),$ i.e.
 \begin{equation}
 \label{ig.eq25}
 \psi = A(\lambda) \cos (\omega t) \quad {\rm and} \quad
 \psi = B(\lambda) \sin (\omega t).
 \end{equation}

 For each function $\psi$ given by (\ref{ig.eq25}) we determine the functions
 $V(t), \ R(t)$ using Eqs. (\ref{ig.eq22}), (\ref{ig.eq23}), (\ref{ig.eq23}), then take the linear
 combinations of the resulting functions  and  arrive at the following
 form of the ``generalized invariant solution" (\ref{ig.eq20}):
  \begin{align}
 & \psi = A(\lambda) \cos (\omega t) + B(\lambda) \sin (\omega t),
 \label{ig.eq26}\\[1ex]
 & v = \frac{fm}{\omega} \left[B'(\lambda) \cos (\omega t) -
 A'(\lambda) \sin (\omega t)\right]  + F (\lambda), \label{ig.eq27}\\[1ex]
 & \rho = \frac{k N^2}{g \omega} \left[B'(\lambda) \cos (\omega t) -
 A'(\lambda) \sin (\omega t)\right]  + H (\lambda), \label{ig.eq28}
 \end{align}
 where $\omega$ is given by Eq. (\ref{ig.eq13}).

 The reckoning shows that the functions
 (\ref{ig.eq26})-(\ref{ig.eq28}) with arbitrary
 $
 A(\lambda), \ B(\lambda)
 $
 solve Eqs. (\ref{iik.eq1})-(\ref{iik.eq3}) provided that $
 F(\lambda), \ H(\lambda)$ satisfy the
 following equation:
 \begin{equation}
 \label{ig.eq29}
 g k H'(\lambda) + f m F'(\lambda) = 0.
 \end{equation}
 One can readily verify that the invariant solution
 (\ref{ig.eq17})-(\ref{ig.eq19}), which is a particular case of
 (\ref{ig.eq26})-(\ref{ig.eq28}), obeys the condition (\ref{ig.eq29}).

 \subsection{Energy of the generalized invariant solution}

 If we substitute in (\ref{ii.ener-dens}) the generalized
 invariant solution (see Eqs. (\ref{ig.eq26})-(\ref{ig.eq28}))
  \begin{align}
 & \psi = A(\lambda) \cos (\omega t) + B(\lambda) \sin (\omega t),
 \label{ig.eq26a}\\[1ex]
 & v = \frac{fm}{\omega} \left[B'(\lambda) \cos (\omega t) -
 A'(\lambda) \sin (\omega t)\right], \label{ig.eq27a}\\[1ex]
 & \rho = \frac{k N^2}{g \omega} \left[B'(\lambda) \cos (\omega t) -
 A'(\lambda) \sin (\omega t)\right], \label{ig.eq28a}
 \end{align}
 where
 $$
 \lambda = kx + m z, \quad \omega^2 = \frac{k^2 N^2 + m^2f^2}{k^2 + m^2}\,,
 $$
 we obtain:
 $$
 E = (k^2 + m^2) \big[A'\,^2 (\lambda) + B'\,^2 (\lambda)\big].
 $$
 Invoking that any conserved vector is defined up to multiplication by an
 arbitrary constant, we divide the above expression for $E$ by
 $(k^2 + m^2)$ and obtain the following energy:
 \begin{equation}
 \label{iik.cl_dil:eq4}
 E = A'\,^2 (\lambda) + B'\,^2 (\lambda).
 \end{equation}

Since the energy density (\ref{iik.cl_dil:eq4}) depends only on
$\lambda = k x + mz,$
 it is constant along the straight line
 \begin{equation}
 \label{iik.cl_dil:eq6}
 k x + m z = {\rm const.}
 \end{equation}
  Accordingly, the ``local energy" (\ref{iik.cl_dil:eq4}) has one and the
  same value at points $(x_0, z_0)$ and $(x_1, z_1)$ provided that
 \begin{equation}
 \label{iik.cl_dil:eq7}
 k x_0 + m z_0 = k x_1 + m z_1.
 \end{equation}

 The energy density (\ref{iik.cl_dil:eq4}) describes the local
 behavior of the solutions. Therefore it is significant to
 understand its distribution on the $(x, z)$ plane.
 Suppose that the functions $A(\lambda), \ B(\lambda)$ and
 their derivatives rapidly decrease as $\eta \rightarrow \infty.$
 If we take, as an example, the functions
 \begin{equation}
 \label{iik.cl_dil:eq5}
 A(\lambda) = \frac{a}{1 + \lambda^2}\,, \quad
 B(\lambda) = \frac{a \lambda}{1 + \lambda^2}\,,
 \end{equation}
 where $a$ is a positive constant, then the energy density (\ref{iik.cl_dil:eq4}) of the
 wave beams has the form
 $$
 E = \frac{a^2}{(1 + \lambda^2)^2}\,\cdot
 $$
 Hence, the energy is localized along  the straight
 line (\ref{iik.cl_dil:eq7}). Therefore
we can define a \textit{wave beam through a point} $(x_0, z_0)$
\textit{as the
 totality of the points} $(x_1, z_1)$ \textit{satisfying Eq.}
 (\ref{iik.cl_dil:eq7}), i.e. identify it with the straight line
 (\ref{iik.cl_dil:eq7}).\\[4ex]
 \null \hfill 6 February 2009

 \end{document}